\documentclass[12pt,a4paper]{article}
\begin{document}
\large

\newpage
\begin{center}
{\bf ON A LATENT STRUCTURE OF LEPTON UNIVERSALITY}
\end{center}
\vspace{1cm}
\begin{center}
{\bf R.S. Sharafiddinov}
\end{center}
\vspace{1cm}
\begin{center}
{\bf Institute of Nuclear Physics, Uzbekistan Academy of Sciences,
Tashkent, 100214 Ulugbek, Uzbekistan}
\end{center}
\vspace{1cm}

The mass of an electroweakly charged lepton consists of the two components
of electric and weak nature, and regardless of the difference in masses, all
leptons have an equal charge, the same charge radius as well as an identical
magnetic moment. Between these currents there appear the most diverse
connections, for example, in their interactions with an electroweak field of
spinless nuclei. We derive the united equations which relate the structural
parts of mass to charge, charge radius and magnetic moment of each lepton as
a consequence of the ideas of flavor symmetry laws. Therefore, these ideas
require the verification of the lepton universality from the point of view
of a constancy of the size implied from the multiplication of a weak mass of
lepton by its electric mass. Such a principle gives the possibility to define
the lepton weak masses. If this picture does not change, leptons universally
interact not only with the photon or the weak boson but also with any of the
corresponding types of gauge fields.

\newpage
\begin{center}
{\bf 1 Introduction}
\end{center}

It is indisputable that naturally united regularities of the nature of
elementary particles establish a flavor independent symmetry between the
interactions of all types of leptons with each of the corresponding types
of gauge bosons. However, over fifty years ago, when the hypothesis of
lepton universality [1-4] was formulated for the first time and now,
when a large number of works \cite{5,6} is dedicated to its different
aspects \cite{7,8}, the question as to what is the unified theoretical
description of this highly astounding symmetry of the micro-world steel has
no unequivocal answer. An exact violation of the lepton universality \cite{9}
is not observed experimentally \cite{10} and the impression that nature itself
testifies in favor of such a symmetrical structure remains. Therefore, to
elucidate the true character of the identical interactions of leptons, it
is desirable not only to use the difference in their masses as a stimulus
to deciding the problem at the new level but also to raise the question as
to whether there exists any mass type dependence of lepton universality and,
if so, what the expected connection says about the mechanism of the discussed
universal phenomena.

Many authors state that there is no rest mass dependence of particle
fundamental interactions and fields. Its existence would seem to contradict
our observation that the electron, muon and tau lepton which possess the same
charge and equal magnetic moment, have difference masses.

It is interesting, however, that the same charge of Coulomb nature may not
be both an electric and a weak charge. At the same time, each of the existing
types of leptons can interact with all of the corresponding types
of gauge fields.

The basis for our approach is that in the framework of the classical theory of
an extensive electron \cite{11,12}, a particle mass has purely electromagnetic
nature. If we accept this idea, the difference in fermion masses would lead
us to the implication that the muon and tau lepton possess some interactions
which are absent in the electron \cite{13}. But the question about their
existence thus far remains open.

On the other hand, as known, the steadiness of the electric charge
distribution in leptons can be explained by the intralepton interratio
of forces of different nature. According to our description \cite{14}, this
implies that to any type of charge there corresponds a kind of mass. Such a
principle reflects the characteristic features of a latent connection between
the mass of a particle and its united field of emission.

At the same time, it is clear that each Dirac particle of Coulomb nature must
have electric \cite{11} as well as weak \cite{15} mass. In other words, mass
$m_{l}$ and charge $e_{l}$ of an electroweakly charged lepton $(l=e,$ $\mu,$
$\tau, ...)$ are equal to all its electroweak $(EW)$ mass and charge
\begin{equation}
m_{l}=m_{l}^{EW}=m_{l}^{E}+m_{l}^{W},
\label{1}
\end{equation}
\begin{equation}
e_{l}=e_{l}^{EW}=e_{l}^{E}+e_{l}^{W},
\label{2}
\end{equation}
consisting of the electric $(E)$ and weak $(W)$ parts. They have a crucial
value for the photon and weak boson leptonic currents.

In conformity with the ideas of quantum electrodynamics, the first of
these currents include the following interaction \cite{13,16} terms:
\begin{equation}
j_{\mu}^{\gamma}=\overline{u}(p',s')[\gamma_{\mu}F_{1l}(q^{2})-
i\sigma_{\mu\lambda}q_{\lambda}F_{2l}(q^{2})]u(p,s),
\label{3}
\end{equation}
where $\sigma_{\mu\lambda}=[\gamma_{\mu},\gamma_{\lambda}]/2,$ $p(s)$
and $p'(s')$ denote the four-momentum (helicities) of incoming and outgoing
fermions, $q=p-p',$ the form factors $F_{il}(q^{2})$ constitute the vector
$V_{l}$ current Dirac $(i=1)$ and Pauli $(i=2)$ components. Here we present
them in the form \cite{17}
\begin{equation}
F_{il}(q^{2})=f_{il}(0)+R_{il}(q^{2})+... .
\label{4}
\end{equation}

The values of $f_{il}(0)$ define the static sizes of the lepton electric
charge and magnetic moment. The terms $R_{il}(q^{2})$ characterize the
particle electromagnetic radius dependence of the form factors.

Here $R_{1l}(q^{2})$ describe the interaction between the charge $r_{l}$
radius of the lepton and the field of emission of the photon:
$R_{1l}(q^{2})=(q^{2}/6)<r^{2}_{l}>.$

It is clear, however, that each type of charge of any lepton constitutes a
kind of leptonic current. According to the standard electroweak theory [18-20],
the latter implies that the terms $f_{1l}$ and $f_{2l}$ describing the Coulomb
scattering of leptons do not coincide with the corresponding sizes from those
form factors which arise in their interactions with the interference field
of emission of the photon and the weak boson \cite{21}.

A given circumstance may serve as a certain indication to an explicit mass
structure dependence of lepton universality. To solve this question one must
refer to the processes on target nuclei, because they can shed light on the
nature of the identical lepton interactions.

Our present work is dedicated to the definition of a latent structure
of the universal interactions of leptons with the Coulomb, the weak and the
electroweak interference fields of emission and of a role in their formation
of the structural components of an electroweak mass. For this purpose
we investigate here the behavior of all types of leptons in the elastic
scattering on a spinless nucleus as a consequence of the availability of
the Coulomb $m_{l}^{E}$ and weak $m_{l}^{W}$ masses and of the electric
$f_{1l}(0)$ charge, charge $r_{l}$ radius and magnetic $f_{2l}(0)$ moment
of longitudinal polarized fermions with vector weak $V_{l}$ currents.

As far as the axial-vector $A_{l}$ currents are concerned, we will start from
the requirement \cite{22} that the same particle possesses simultaneously only
one of the currents, $V_{l}$ or $A_{l}$ but not both of them. Our reasoning
refers to those leptons, among which there are no fermions with $A_{l}$
currents.

\begin{center}
{\bf 2 Unity of lepton vector electroweak interaction structural parts}
\end{center}

The scattering of leptons by nuclei in the limit of one-boson exchange
can be described by the matrix elements \cite{21}
$$M^{E}_{fi}=\frac{4\pi\alpha}{q_{E}^{2}}\overline{u}(p_{E}',s')
\{\gamma_{\mu}[f_{1l}^{E}(0)+\frac{1}{6}q_{E}^{2}<r^{2}_{l}>_{E}]-$$
\begin{equation}
-i\sigma_{\mu\lambda}q_{\lambda E}f_{2l}^{E}(0)\}
u(p_{E},s)<f|J_{\mu}^{\gamma}(q_{E})|i>,
\label{5}
\end{equation}
\begin{equation}
M^{W}_{fi}=
\frac{G_{F}}{\sqrt{2}}\overline{u}(p_{W}',s')\gamma_{\mu}
g_{V_{l}}^{*}u(p_{W},s)<f|J_{\mu}^{Z}(q_{W})|i>.
\label{6}
\end{equation}
Here $l=l_{L,R}^{-}(l_{R,L}^{+}),$ $q_{E}=p_{E}-p_{E}',$ $q_{W}=p_{W}-p_{W}',$
$p_{E}(p_{W})$ and $p_{E}'(p_{W}')$ imply the four-momentum of a particle
before and after the Coulomb (weak) interaction, the functions $f_{il}^{E}$
and $<r^{2}_{l}>_{E}$ describe a latent electric $m_{l}^{E}$
mass dependence of the lepton charge, charge radius and magnetic moment,
$J_{\mu}^{\gamma}$ and $J_{\mu}^{Z}$ characterize the nuclear currents in
the processes with photon and $Z$-boson, $g_{V_{l}}^{*}$ distinguishes from
$g_{V_{l}},$ namely, from the weak leptonic current vector part constant by a
multiplier $(1/\sin\theta_{W})$ which appears for the case $e_{l}^{E}=1$ when
\begin{equation}
e_{l}^{E}=e_{l}^{W}\sin\theta_{W}.
\label{7}
\end{equation}

We have already mentioned that any type of lepton can have simultaneously
both Coulomb and weak masses \cite{11,12,15}. Therefore, from the point of
view of each of the scattering amplitude (\ref{5}) or (\ref{6}), it should
be added \cite{21} that to the process originating at the expense of the
exchange by the two bosons of Coulomb and weak nature, responds the mixed
interference $(I)$ interaction
$$ReM^{E}_{fi}M^{*W}_{fi}=
\frac{4\pi\alpha G_{F}}{\sqrt{2}q_{I}^{2}}
Re\Lambda_{I}\Lambda_{I}'\{\gamma_{\mu}
[f_{1l}^{I}(0)+$$
$$+\frac{1}{6}q_{I}^{2}<r^{2}_{l}>_{I}]-$$
\begin{equation}
-i\sigma_{\mu\lambda}q_{\lambda I}f_{2l}^{I}(0)\}
\gamma_{\mu}g_{V_{l}}^{*}J_{\mu}^{\gamma}(q_{I})J_{\mu}^{Z}(q_{I}).
\label{8}
\end{equation}

The appearance of $f_{il}^{I}$ and $<r^{2}_{l}>_{I}$ in it is explained
by the availability of a compound structure of mass and charge. We have
also used the sizes
$$q_{I}=p_{I}-p_{I}',$$
$$\Lambda_{I}=u(p_{I},s)\overline{u}(p_{I},s),$$
$$\Lambda_{I}'=u(p_{I}',s')\overline{u}(p_{I}',s').$$
Here $p_{I}$ and $p_{I}'$ express the four-momentum of the lepton before
and after the electroweak interference emission.

Such a presentation can be based on the fact that the interference $m_{l}^{I}$
mass of the lepton does not coincide with all its $m_{l}^{EW}$ rest mass. This
distinction between the values of $m_{l}^{I}$ and $m_{l}^{EW}$ appears in the
difference of the electric $m_{l}^{E}$ and weak $m_{l}^{W}$ components of 
the lepton mass. Their connection similarly to relation (\ref{7}) will be
responsible for the electroweak unification at a more fundamental
dynamical level.

In the case of nuclei with the electric $(Z)$ and weak $(Z_{W})$ charges and
of leptons of longitudinal polarization, the differential cross-section
of the studied process on the basis of (\ref{5})-(\ref{8}) and
the standard definition
\begin{equation}
\frac{d\sigma_{EW}(s,s')}{d\Omega}=
\frac{1}{16\pi^{2}}|M^{E}_{fi}+M^{W}_{fi}|^{2}
\label{9}
\end{equation}
can be presented in the following manner:
$$d\sigma_{EW}^{V_{l}}(\theta_{EW},s,s')=
d\sigma_{E}^{V_{l}}(\theta_{E},s,s')+$$
\begin{equation}
+d\sigma_{I}^{V_{l}}(\theta_{I},s,s')+
d\sigma_{W}^{V_{l}}(\theta_{W},s,s'),
\label{10}
\end{equation}
where $\theta_{EW}$ is the scattering angle of lepton after its interaction
with an electroweakly united $(EW)$ field of emission.

To the first term in (\ref{10}) there corresponds the Coulomb scattering
and has the form
$$\frac{d\sigma_{E}^{V_{l}}(\theta_{E},s,s')}{d\Omega}=
\frac{1}{2}\sigma^{E}_{o}(1-\eta^{2}_{E})^{-1}\{(1+ss')[f_{1l}^{E}-$$
$$-\frac{2}{3}<r^{2}_{l}>_{E}(m_{l}^{E})^{2}\gamma_{E}^{-1}]^{2}+$$
$$+\eta^{2}_{E}(1-ss')[(f_{1l}^{E}-
\frac{2}{3}<r^{2}_{l}>_{E}(m_{l}^{E})^{2}\gamma_{E}^{-1})^{2}+$$
\begin{equation}
+4(m_{l}^{E})^{2}(1-\eta^{-2}_{E})^{2}(f_{2l}^{E})^{2}]
tg^{2}\frac{\theta_{E}}{2}\}
F_{E}^{2}(q_{E}^{2}).
\label{11}
\end{equation}

The interference cross-section explained by the electroweakly mixed
interaction (\ref{8}) is equal to
$$\frac{d\sigma_{I}^{V_{l}}(\theta_{I},s,s')}{d\Omega}=
\frac{1}{2}\rho_{I}\sigma^{I}_{o}(1-\eta^{2}_{I})^{-1}
g_{V_{l}}\{(1+ss')[f_{1l}^{I}-$$
$$-\frac{2}{3}<r^{2}_{l}>_{I}(m_{l}^{I})^{2}\gamma_{I}^{-1}]+
\eta^{2}_{I}(1-ss')[f_{1l}^{I}-$$
\begin{equation}
-\frac{2}{3}<r^{2}_{l}>_{I}(m_{l}^{I})^{2}\gamma_{I}^{-1}]
tg^{2}\frac{\theta_{I}}{2}\}F_{I}(q_{I}^{2}).
\label{12}
\end{equation}

The purely weak contributions are written as
$$\frac{d\sigma_{W}^{V_{l}}(\theta_{W},s,s')}{d\Omega}=
\frac{G_{F}^{2}(m_{l}^{W})^{2}}{16\pi^{2}\sin^{2}\theta_{W}}g_{V_{l}}^{2}
\{\eta_{W}^{-2}(1+ss')\cos^{2}\frac{\theta_{W}}{2}+$$
\begin{equation}
+(1-ss')\sin^{2}\frac{\theta_{W}}{2}\}F_{W}^{2}(q_{W}^{2}).
\label{13}
\end{equation}
Here we must keep in mind that
$$\sigma_{o}^{E}=\frac{\alpha^{2}}{4(m_{l}^{E})^{2}}
\frac{\gamma_{E}^{2}}{\alpha_{E}}, \, \, \, \,
\rho_{I}=-\frac{2G_{F}(m_{l}^{I})^{2}}
{\pi\sqrt{2}\alpha\sin\theta_{W}}\gamma_{I}^{-1},$$
$$\sigma_{o}^{I}=\frac{\alpha^{2}}{4(m_{l}^{I})^{2}}
\frac{\gamma_{I}^{2}}{\alpha_{I}}, \, \, \, \,
\alpha_{K}=\frac{\eta^{2}_{K}}{(1-\eta^{2}_{K})
\cos^{2}(\theta_{K}/2)},$$
$$\gamma_{K}=\frac{\eta^{2}_{K}}{(1-\eta^{2}_{K})
\sin^{2}(\theta_{K}/2)}, \, \, \, \,
\eta_{K}=\frac{m_{l}^{K}}{E_{l}^{K}},$$
$$F_{E}(q_{E}^{2})=ZF_{c}(q_{E}^{2}), \, \, \, \,
F_{I}(q_{I}^{2})=ZZ_{W}F_{c}^{2}(q_{I}^{2}),$$
$$F_{W}(q_{W}^{2})=Z_{W}F_{c}(q_{W}^{2}), \, \, \, \,
q_{K}^{2}=-4(m_{l}^{K})^{2}\gamma_{K}^{-1},$$
$$Z_{W}=\frac{1}{2}\{\beta_{V}^{(0)}(Z+N)+\beta_{V}^{(1)}(Z-N)\},$$
$$A=Z+N, \, \, \, \, M_{T}=\frac{1}{2}(Z-N),$$
$$\beta_{V}^{(0)}=-2\sin^{2}\theta_{W},
\, \, \, \, \beta_{V}^{(1)}=\frac{1}{2}-2\sin^{2}\theta_{W},$$
$$g_{V_{l}}=-\frac{1}{2}+2\sin^{2}\theta_{W}, \, \, \, \, K=E,I,W.$$

Among them $\theta_{K}$ are the scattering angles in the Coulomb,
interference and purely weak processes at the energies of leptons $E_{l}^{K},$
the functions $F_{c}(q_{K}^{2})$ are the charge $(F_{c}(0)=1)$ form factors of
a nucleus in these three types of interactions, $M_{T}$ is the projection of
its isospin $T,$ $\beta_{V}^{(0)}$ and $\beta_{V}^{(1)}$ are the isoscalar
and isovector constants of the nuclear vector weak current.

Formula (\ref{11}) contains the purely self-interference contributions
$(f_{il}^{E})^{2}$ and $<r_{l}^{4}>_{E}$ as well as the contribution
$f_{1l}^{E}<r_{l}^{2}>_{E}$ of the mixed interference between the interactions
of the lepton charge and charge radius with the photon. Any of these components
of the Coulomb scattering cross-section responds to the formation of the
left - or right-handed \cite{23,24} dileptons of the vector currents
\begin{equation}
(l^{-}_{L}, l^{+}_{R}), \, \, \, \, (l^{-}_{R}, l^{+}_{L}).
\label{14}
\end{equation}

With the availability of the interaction (\ref{8}), each of such parafermions
must lead to the appearance in the scattering cross-section (\ref{12}) of one
of its structural parts $g_{V_{l}}f_{1l}^{I}$ or $g_{V_{l}}<r_{l}^{2}>_{I}$
and that, consequently, the behavior of dileptons in the nuclear charge field
depends on the nature of their interaction. Therefore, from the point of view
of (\ref{13}) and its self-interference contributions $g_{V_{l}}^{2},$ it
should be expected that the possibility for the existence of parafermions
of the weak currents is not excluded.

Turning to (\ref{9}), we remark that it redoubles the size of the mixed
interference cross-sections. However, according to the considerations of
symmetry, the number of dileptons and the structural phenomena in which
they appear coincide. This conformity requires the separation of any type
of the mixed interference contribution into the two equal parts.

Furthermore, if it turns out that the terms $(1+ss')$ and $(1-ss')$ in each
of (\ref{11})-(\ref{13}) impliy the existence in left $(s=-1)$ - and right
$(s=+1)$-handed leptons of vector nature of those types of interactions
which are responsible for the scattering with $(s'=s)$ or without $(s'=-s)$
flip of their helicities, taking into account of the latter, one can replace
eq. (\ref{10}) with
$$d\sigma_{EW}^{V_{l}}(\theta_{EW},s)=d\sigma_{E}^{V_{l}}(\theta_{E},s)+
\frac{1}{2}d\sigma_{I}^{V_{l}}(\theta_{I},s)+$$
\begin{equation}
+\frac{1}{2}d\sigma_{I}^{V_{l}}(\theta_{I},s)+
d\sigma_{W}^{V_{l}}(\theta_{W},s),
\label{15}
\end{equation}
where to the purely Coulomb scattering there correspond the cross sections
$$d\sigma_{E}^{V_{l}}(\theta_{E},s)=
d\sigma_{E}^{V_{l}}(\theta_{E},f_{1l}^{E},s)+
\frac{1}{2}d\sigma_{E}^{V_{l}}(\theta_{E},f_{1l}^{E},<r^{2}_{l}>_{E},s)+$$
$$+\frac{1}{2}d\sigma_{E}^{V_{l}}(\theta_{E},f_{1l}^{E},<r^{2}_{l}>_{E},s)+$$
\begin{equation}
+d\sigma_{E}^{V_{l}}(\theta_{E},<r^{2}_{l}>_{E},s)+
d\sigma_{E}^{V_{l}}(\theta_{E},f_{2l}^{E},s),
\label{16}
\end{equation}

$$\frac{d\sigma_{E}^{V_{l}}(\theta_{E},f_{1l}^{E},s)}{d\Omega}=
\frac{d\sigma_{E}^{V_{l}}(\theta_{E},f_{1l}^{E},s'=s)}{d\Omega}+
\frac{d\sigma_{E}^{V_{l}}(\theta_{E},f_{1l}^{E},s'=-s)}{d\Omega}=$$
\begin{equation}
=\sigma^{E}_{o}(1-\eta^{2}_{E})^{-1}(1+\eta_{E}^{2}tg^{2}
\frac{\theta_{E}}{2})(f_{1l}^{E})^{2}F_{E}^{2}(q^{2}_{E}),
\label{17}
\end{equation}

$$\frac{d\sigma_{E}^{V_{l}}(\theta_{E},f_{1l}^{E},<r^{2}_{l}>_{E},s)}
{d\Omega}=
\frac{d\sigma_{E}^{V_{l}}(\theta_{E},f_{1l}^{E},<r^{2}_{l}>_{E},s'=s)}
{d\Omega}+$$
$$+\frac{d\sigma_{E}^{V_{l}}(\theta_{E},f_{1l}^{E},<r^{2}_{l}>_{E},s'=-s)}
{d\Omega}=$$
$$=-\frac{2}{3}(m_{l}^{E})^{2}\gamma_{E}^{-1}\sigma^{E}_{o}
(1-\eta^{2}_{E})^{-1}\times $$
\begin{equation}
\times (1+\eta_{E}^{2}tg^{2}\frac{\theta_{E}}{2})
f_{1l}^{E}<r^{2}_{l}>_{E}F_{E}^{2}(q^{2}_{E}),
\label{18}
\end{equation}

$$\frac{d\sigma_{E}^{V_{l}}(\theta_{E},<r^{2}_{l}>_{E},s)}{d\Omega}=
\frac{d\sigma_{E}^{V_{l}}(\theta_{E},<r^{2}_{l}>_{E},s'=s)}{d\Omega}+$$
$$+\frac{d\sigma_{E}^{V_{l}}(\theta_{E},<r^{2}_{l}>_{E},s'=-s)}{d\Omega}=$$
$$=\frac{4}{9}(m_{l}^{E})^{4}\gamma_{E}^{-2}\sigma^{E}_{o}
(1-\eta^{2}_{E})^{-1}\times $$
\begin{equation}
\times (1+\eta_{E}^{2}tg^{2}\frac{\theta_{E}}{2})
<r^{4}_{l}>_{E}F_{E}^{2}(q^{2}_{E}),
\label{19}
\end{equation}

$$\frac{d\sigma_{E}^{V_{l}}(\theta_{E},f_{2l}^{E},s)}{d\Omega}=
\frac{d\sigma_{E}^{V_{l}}(\theta_{E},f_{2l}^{E},s'=-s)}{d\Omega}=$$
\begin{equation}
=4(m_{l}^{E})^{2}\eta^{-2}_{E}\sigma^{E}_{o}(1-\eta^{2}_{E})^{2}
(f_{2l}^{E})^{2}F_{E}^{2}(q^{2}_{E})tg^{2}\frac{\theta_{E}}{2}.
\label{20}
\end{equation}

The second term in (\ref{15}) responds to the scattering originating at the
expense of the electroweak interference and must have the following structure:
$$d\sigma_{I}^{V_{l}}(\theta_{I},s)=
d\sigma_{I}^{V_{l}}(\theta_{I},g_{V_{l}},f_{1l}^{I},s)+$$
\begin{equation}
+d\sigma_{I}^{V_{l}}(\theta_{I},g_{V_{l}},<r^{2}_{l}>_{I},s),
\label{21}
\end{equation}

$$\frac{d\sigma_{I}^{V_{l}}(\theta_{I},g_{V_{l}},f_{1l}^{I},s)}{d\Omega}=
\frac{d\sigma_{I}^{V_{l}}(\theta_{I},g_{V_{l}},f_{1l}^{I},s'=s)}
{d\Omega}+$$
$$+\frac{d\sigma_{I}^{V_{l}}(\theta_{I},g_{V_{l}},f_{1l}^{I},s'=-s)}
{d\Omega}=$$
$$=\rho_{I}\sigma^{I}_{o}(1-\eta^{2}_{I})^{-1}\times $$
\begin{equation}
\times (1+\eta_{I}^{2}tg^{2}\frac{\theta_{I}}{2})
g_{V_{l}}f_{1l}^{I}F_{I}(q_{I}^{2}),
\label{22}
\end{equation}

$$\frac{d\sigma_{I}^{V_{l}}(\theta_{I},g_{V_{l}},<r^{2}_{l}>_{I},s)}
{d\Omega}=
\frac{d\sigma_{I}^{V_{l}}(\theta_{I},g_{V_{l}},<r^{2}_{l}>_{I},s'=s)}
{d\Omega}+$$
$$+\frac{d\sigma_{I}^{V_{l}}(\theta_{I},g_{V_{l}},<r^{2}_{l}>_{I},s'=-s)}
{d\Omega}=$$
$$=-\frac{2}{3}(m_{l}^{I})^{2}\gamma_{I}^{-1}
\rho_{I}\sigma^{I}_{o}(1-\eta^{2}_{I})^{-1}\times $$
\begin{equation}
\times (1+\eta_{I}^{2}tg^{2}\frac{\theta_{I}}{2})
g_{V_{l}}<r^{2}_{l}>_{I}F_{I}(q_{I}^{2}).
\label{23}
\end{equation}

In the same way one can write the cross-section of purely weak interaction
of partially longitudinally polarized leptons
$$\frac{d\sigma_{W}^{V_{l}}(\theta_{W},g_{V_{l}},s)}{d\Omega}=
\frac{d\sigma_{W}^{V_{l}}(\theta_{W},g_{V_{l}},s'=s)}{d\Omega}+
\frac{d\sigma_{W}^{V_{l}}(\theta_{W},g_{V_{l}},s'=-s)}{d\Omega}=$$
\begin{equation}
=\frac{G_{F}^{2}(m_{l}^{W})^{2}}{8\pi^{2}\sin^{2}\theta_{W}}
\eta_{W}^{-2}(1+\eta_{W}^{2}tg^{2}\frac{\theta_{W}}{2})
g_{V_{l}}^{2}F_{W}^{2}(q_{W}^{2})\cos^{2}\frac{\theta_{W}}{2}.
\label{24}
\end{equation}

By following the structure of formulas (\ref{16})-(\ref{24}), it is easy
to observe the negative signs of the cross-sections (\ref{18}) and (\ref{23})
which testify in favor of the coexistence of the electric $f_{1l}(0)$ charge
of lepton and its charge $r_{l}$ radius. On the other hand, as follows from
our earlier developments, between $f_{1l}$ and $f_{2l}$ there exists a
connection \cite{16,25}. At the same time, the lepton itself possesses
simultaneously each of these three types of vector currents.

To establish their nature, it is desirable to reduce (\ref{9}) after averaging
the cross-sections (\ref{11})-(\ref{13}) over $s$ and summing over $s'$
to the form
$$d\sigma_{EW}^{V_{l}}(\theta_{EW})=d\sigma_{E}^{V_{l}}(\theta_{E})+
\frac{1}{2}d\sigma_{I}^{V_{l}}(\theta_{I})+$$
\begin{equation}
+\frac{1}{2}d\sigma_{I}^{V_{l}}(\theta_{I})+
d\sigma_{W}^{V_{l}}(\theta_{W}).
\label{25}
\end{equation}

As well as in (\ref{15}), any term here describes a kind of process and
has the most diverse structure:
$$d\sigma_{E}^{V_{l}}(\theta_{E})=
d\sigma_{E}^{V_{l}}(\theta_{E},f_{1l}^{E})+
\frac{1}{2}d\sigma_{E}^{V_{l}}(\theta_{E},f_{1l}^{E},<r^{2}_{l}>_{E})+$$
$$+\frac{1}{2}d\sigma_{E}^{V_{l}}(\theta_{E},f_{1l}^{E},<r^{2}_{l}>_{E})+$$
\begin{equation}
+d\sigma_{E}^{V_{l}}(\theta_{E},<r^{2}_{l}>_{E})+
d\sigma_{E}^{V_{l}}(\theta_{E},f_{2l}^{E}),
\label{26}
\end{equation}
\begin{equation}
d\sigma_{I}^{V_{l}}(\theta_{I})=
d\sigma_{I}^{V_{l}}(\theta_{I},g_{V_{l}},f_{1l}^{I})+
d\sigma_{I}^{V_{l}}(\theta_{I},g_{V_{l}},<r^{2}_{l}>_{I}),
\label{27}
\end{equation}
\begin{equation}
d\sigma_{W}^{V_{l}}(\theta_{W})=
d\sigma_{W}^{V_{l}}(\theta_{W},g_{V_{l}}).
\label{28}
\end{equation}

Each cross-section in them coincides with the corresponding components from
solutions (\ref{16}), (\ref{21}), (\ref{24}) and, consequently, we have
\begin{equation}
d\sigma_{E}^{V_{l}}(\theta_{E},f_{1l}^{E})=
d\sigma_{E}^{V_{l}}(\theta_{E},f_{1l}^{E},s),
\label{29}
\end{equation}
\begin{equation}
d\sigma_{E}^{V_{l}}(\theta_{E},f_{1l}^{E},<r^{2}_{l}>_{E})=
d\sigma_{E}^{V_{l}}(\theta_{E},f_{1l}^{E},<r^{2}_{l}>_{E},s),
\label{30}
\end{equation}
\begin{equation}
d\sigma_{E}^{V_{l}}(\theta_{E},<r^{2}_{l}>_{E})=
d\sigma_{E}^{V_{l}}(\theta_{E},<r^{2}_{l}>_{E},s),
\label{31}
\end{equation}
\begin{equation}
d\sigma_{E}^{V_{l}}(\theta_{E},f_{2l}^{E})=
d\sigma_{E}^{V_{l}}(\theta_{E},f_{2l}^{E},s),
\label{32}
\end{equation}
\begin{equation}
d\sigma_{I}^{V_{l}}(\theta_{I},g_{V_{l}},f_{1l}^{I})=
d\sigma_{I}^{V_{l}}(\theta_{I},g_{V_{l}},f_{1l}^{I},s),
\label{33}
\end{equation}
\begin{equation}
d\sigma_{I}^{V_{l}}(\theta_{I},g_{V_{l}},<r^{2}_{l}>_{I})=
d\sigma_{I}^{V_{l}}(\theta_{I},g_{V_{l}},<r^{2}_{l}>_{I},s),
\label{34}
\end{equation}
\begin{equation}
d\sigma_{W}^{V_{l}}(\theta_{W},g_{V_{l}})=
d\sigma_{W}^{V_{l}}(\theta_{W},g_{V_{l}},s).
\label{35}
\end{equation}

Basing on (\ref{15}) and (\ref{25}), one can think that either an incoming
lepton flux consists of longitudinal polarized particles or in it there are
no fermions of any spin polarization. It is not difficult to see, however,
that this is not quite so. The point is that the behavior of elementary
particles of a definite helicity depends not only on their dynamical
properties \cite{16,26} but also on the nature of the medium \cite{27} in
which they interact with matter. In other words, a fermion passing through the
field of emission undergoes a strong change in its spin direction. Thereby, the
possibility of the presence of both longitudinally polarized and unpolarized
leptons in fluxes of initial and final fermions is not excluded. The scattered
particles can therefore constitute a partially ordered set of outgoing leptons.

Of course, to this class there corresponds a set of cross-sections. In our
case, from (\ref{15}) and (\ref{25}), we are led to the following class:
\begin{equation}
d\sigma_{EW}^{V_{l}}=\{d\sigma_{EW}^{V_{l}}(\theta_{EW},s), \, \, \, \,
d\sigma_{EW}^{V_{l}}(\theta_{EW})\}.
\label{36}
\end{equation}

Its elements describe a situation in which any of the cross sections
(\ref{15}) and (\ref{25}) constitutes a kind of subset:
$$d\sigma_{EW}^{V_{l}}(\theta_{EW},s)=
\{d\sigma_{E}^{V_{l}}(\theta_{E},f_{1l}^{E},s),  \, \, \, \,
\frac{1}{2}d\sigma_{E}^{V_{l}}(\theta_{E},f_{1l}^{E},<r^{2}_{l}>_{E},s),$$
$$\frac{1}{2}d\sigma_{E}^{V_{l}}(\theta_{E},f_{1l}^{E},<r^{2}_{l}>_{E},s),
\, \, \, \,
d\sigma_{E}^{V_{l}}(\theta_{E},<r^{2}_{l}>_{E},s),$$
$$d\sigma_{E}^{V_{l}}(\theta_{E},f_{2l}^{E},s),
\, \, \, \,
\frac{1}{2}d\sigma_{I}^{V_{l}}(\theta_{I},g_{V_{l}},f_{1l}^{I},s),$$
$$\frac{1}{2}d\sigma_{I}^{V_{l}}(\theta_{I},g_{V_{l}},f_{1l}^{I},s),
\, \, \, \,
\frac{1}{2}d\sigma_{I}^{V_{l}}(\theta_{I},g_{V_{l}},<r^{2}_{l}>_{I},s),$$
\begin{equation}
\frac{1}{2}d\sigma_{I}^{V_{l}}(\theta_{I},g_{V_{l}},<r^{2}_{l}>_{I},s),
\, \, \, \,
d\sigma_{W}^{V_{l}}(\theta_{W},g_{V_{l}},s)\},
\label{37}
\end{equation}
$$d\sigma_{EW}^{V_{l}}(\theta_{EW})=
\{d\sigma_{E}^{V_{l}}(\theta_{E},f_{1l}^{E}),
\, \, \, \,
\frac{1}{2}d\sigma_{E}^{V_{l}}(\theta_{E},f_{1l}^{E},<r^{2}_{l}>_{E}),$$
$$\frac{1}{2}d\sigma_{E}^{V_{l}}(\theta_{E},f_{1l}^{E},<r^{2}_{l}>_{E}),
\, \, \, \,
d\sigma_{E}^{V_{l}}(\theta_{E},<r^{2}_{l}>_{E}),$$
$$d\sigma_{E}^{V_{l}}(\theta_{E},f_{2l}^{E}), \, \, \, \,
\frac{1}{2}d\sigma_{I}^{V_{l}}(\theta_{I},g_{V_{l}},f_{1l}^{I}),$$
$$\frac{1}{2}d\sigma_{I}^{V_{l}}(\theta_{I},g_{V_{l}},f_{1l}^{I}),
\, \, \, \,
\frac{1}{2}d\sigma_{I}^{V_{l}}(\theta_{I},g_{V_{l}},<r^{2}_{l}>_{I}),$$
\begin{equation}
\frac{1}{2}d\sigma_{I}^{V_{l}}(\theta_{I},g_{V_{l}},<r^{2}_{l}>_{I}),
\, \, \, \,
d\sigma_{W}^{V_{l}}(\theta_{W},g_{V_{l}})\}.
\label{38}
\end{equation}

Both subclasses, according to (\ref{29})-(\ref{35}), must be equal. However,
a coincidence of cross-sections (\ref{15}) and (\ref{25}) can happen only
in the case where they respond to the different processes of formation of
the same dileptons. Such a possibility is realized if to all elements of
sets (\ref{37}) and (\ref{38}) there corresponds one of the two types of
spin states of parafermions. This speaks in favor of the equality of the
studied process cross-section structural parts.

Another serious basis for such a universality is the existence in leptonic
families of doublets of flavor symmetrical connection \cite{22} of the two
left (right)-handed leptons of a definite type. A similar dependence can
appear as a consequence of the availability of the unified force responsible
for the structure of dileptons. This unification requires the use the flavor
symmetry as a theorem \cite{28,29} about the equality of cross-sections of the
interaction with the field of emission of lepton vector electroweak current
structural components.

\begin{center}
{\bf 3 Relation of lepton vector currents of definite flavor}
\end{center}

We see that between the currents $f_{1l}^{E},$ $<r_{l}^{2}>_{E},$ $f_{2l}^{E},$ 
$f_{1l}^{I},$ $<r_{l}^{2}>_{I}$ and $g_{V_{l}}$ there exist hard connections, 
owing to which the interrelationship of each pair of elements in (\ref{36}) 
is not different from the unity. Thereby, this allows to establish 
the forty-two relations.

Together with the sizes of cross-sections (\ref{29})-(\ref{35}), the latter
constitute a system of the twenty-one most diverse equations. They are of
course connected with the functions of the variables $E_{l}^{K},$ $\eta_{K},$
$\theta_{K},$ and thus directly with the square of the momentum transfer.
Therefore, for the establishment of their explicit form, it should be
chosen $E_{l}^{K}$ so that to the case $q_{K}^{2}\rightarrow 0$
there corresponds a kind of scattering angle.

At large energies $(E_{l}^{K}\gg m_{l}^{K})$ when $\eta_{K}\rightarrow 0,$
$q_{K}^{2}\rightarrow 0,$ $\theta_{K}\rightarrow 0,$ the exploring limits
are reduced to the exact values \cite{22,30}, because of which the structure
of the investigated relations becomes fully definite.

To elucidate these ideas, it is desirable to use five equations from
the starting system:
\begin{equation}
\frac{d\sigma_{E}^{V_{l}}(\theta_{E},<r^{2}_{l}>_{E})}
{d\sigma_{E}^{V_{l}}(\theta_{E},f_{il}^{E})}=1,
\, \, \, \,
\frac{d\sigma_{I}^{V_{l}}(\theta_{I},g_{V_{l}},f_{1l}^{I})}
{2d\sigma_{E}^{V_{l}}(\theta_{E},f_{1l}^{E})}=1,
\label{39}
\end{equation}
\begin{equation}
\frac{2d\sigma_{W}^{V_{l}}(\theta_{W},g_{V_{l}})}
{d\sigma_{E}^{V_{l}}(\theta_{E},f_{1l}^{E},<r^{2}_{l}>_{E})}=1,
\, \, \, \,
\frac{d\sigma_{I}^{V_{l}}(\theta_{I},g_{V_{l}},<r^{2}_{l}>_{I})}
{d\sigma_{I}^{V_{l}}(\theta_{I},g_{V_{l}},f_{1l}^{I})}=1.
\label{40}
\end{equation}

Uniting (\ref{39}) and (\ref{40}) with the expressions of the cross-sections
(\ref{29})-(\ref{35}) and having in mind the limits
$$lim_{\eta_{K}\rightarrow 0,\theta_{K}\rightarrow 0}
\frac{\eta^{2}_{K}}{(1-\eta^{2}_{K})
\sin^{2}(\theta_{K}/2)}=-2,$$
$$lim_{\eta_{E}\rightarrow 0,\theta_{E}\rightarrow 0}
\frac{\eta^{2}_{E}\sin^{-2}(\theta_{E}/2)}
{(1+\eta_{E}^{2}tg^{2}(\theta_{E}/2))
\cos^{2}(\theta_{E}/2)}=4,$$
one can find again that
\begin{equation}
\frac{1}{3}<r^{2}_{l}>_{E}(m_{l}^{E})^{2}=\pm f_{1l}^{E}(0),
\label{41}
\end{equation}
\begin{equation}
\frac{1}{6}<r^{2}_{l}>_{E}m_{l}^{E}=\pm f_{2l}^{E}(0),
\label{42}
\end{equation}
\begin{equation}
(f_{1l}^{E}(0))^{2}=g_{V_{l}}\frac{G_{F}(m_{l}^{E})^{2}}
{2\pi\sqrt{2}\alpha}\frac{Z_{W}^{*}}{\sin\theta_{W}}f_{1l}^{I}(0),
\label{43}
\end{equation}
\begin{equation}
\frac{1}{3}<r^{2}_{l}>_{E}f_{1l}^{E}(0)=g_{V_{l}}^{2}
\frac{G_{F}^{2}(m_{l}^{W})^{2}}{8\pi^{2}\alpha^{2}}
\left(\frac{Z_{W}^{*}}{\sin\theta_{W}}\right)^{2},
\label{44}
\end{equation}
\begin{equation}
\frac{1}{3}<r^{2}_{l}>_{I}(m_{l}^{I})^{2}=f_{1l}^{I}(0).
\label{45}
\end{equation}

The presence of $Z_{W}^{*}=Z_{W}/Z$ in (\ref{43}) and (\ref{44}) simultaneously
includes in the discussion the possible change of self-values of $f_{il}^{E},$
$<r^{2}_{l}>_{E},$ $f_{1l}^{I}$ and $<r^{2}_{l}>_{I}$ in the target nucleus
isotopic structure dependence \cite{31}. Such a regularity, however, encounters
many problems which give the possibility to raise the most diverse questions
of fundamental importance. Making all of them explicit would take too long
since they should be described in detail, requiring a lengthy presentation
in terms of number of pages.

\begin{center}
{\bf 4 Mass structure of vector leptons}
\end{center}

It is already clear from the above reasoning that the target nucleus isotopic
structure can essentially change the general picture of elastic scattering.
Therefore, to exclude this influence, we must at first choose a nucleus
with an equal number of neutrons and protons.

Solving eqs. (\ref{41})-(\ref{45}) at $N=Z,$ we establish here an explicit
mass structure dependence of leptonic currents of vector nature
\begin{equation}
f_{1l}^{E}(0)=-g_{V_{l}}
\frac{G_{F}m_{l}^{E}m_{l}^{W}}{\pi\sqrt{2}\alpha}\sin\theta_{W},
\label{46}
\end{equation}
\begin{equation}
f_{2l}^{E}(0)=-g_{V_{l}}
\frac{G_{F}m_{l}^{W}}{2\pi\sqrt{2}\alpha}\sin\theta_{W},
\label{47}
\end{equation}
\begin{equation}
<r^{2}_{l}>_{E}=-g_{V_{l}}
\frac{3G_{F}}{\pi\sqrt{2}\alpha}\frac{m_{l}^{W}}{m_{l}^{E}}\sin\theta_{W},
\label{48}
\end{equation}
\begin{equation}
f_{1l}^{I}(0)=-g_{V_{l}}
\frac{G_{F}(m_{l}^{W})^{2}}{\pi\sqrt{2}\alpha}\sin\theta_{W},
\label{49}
\end{equation}
\begin{equation}
<r^{2}_{l}>_{I}=-g_{V_{l}}\frac{3G_{F}}{\pi\sqrt{2}\alpha}
\left(\frac{m_{l}^{W}}{m_{l}^{I}}\right)^{2}\sin\theta_{W}.
\label{50}
\end{equation}

Taking into account that to
\begin{equation}
e_{l}^{E}=-g_{V_{l}}
\frac{G_{F}m_{l}^{E}m_{l}^{W}}{\pi\sqrt{2}\alpha}\sin\theta_{W}
\label{51}
\end{equation}
there corresponds the vector lepton renormalized electric charge, it is not
difficult to express (\ref{46})-(\ref{50}) in a latent united form:
\begin{equation}
f_{1l}^{E}(0)=e_{l}^{E},
\label{52}
\end{equation}
\begin{equation}
f_{2l}^{E}(0)=\frac{e_{l}^{E}}{2m_{l}^{E}},
\label{53}
\end{equation}
\begin{equation}
<r^{2}_{l}>_{E}=\frac{3e_{l}^{E}}{(m_{l}^{E})^{2}},
\label{54}
\end{equation}
\begin{equation}
f_{1l}^{I}(0)=\frac{m_{l}^{W}}{m_{l}^{E}}e_{l}^{E},
\label{55}
\end{equation}
\begin{equation}
<r^{2}_{l}>_{I}=\frac{m_{l}^{E}m_{l}^{W}}{(m_{l}^{I})^{2}}<r^{2}_{l}>_{E}.
\label{56}
\end{equation}

So, we have learned that $f_{2l}^{E}$ coincides with the Dirac value of
the magnetic moment. As far as its anomalous part \cite{32} is concerned,
it can appear only in the case when the interaction originates at the expense
of the exchange by the two photons.

\begin{center}
{\bf 5 Conclusion}
\end{center}

If we suppose \cite{33} in conclusion that $\sin^{2}\theta_{W}=0.231,$
jointly with well-known laboratory facts \cite{34}, the solution (\ref{51})
predicts the sizes of the lepton weak masses
$m_{e}^{W}=5.15\cdot 10^{-2}\ {\rm eV},$
$m_{\mu}^{W}=2.49\cdot 10^{-4}\ {\rm eV},$
$m_{\tau}^{W}=1.48\cdot 10^{-5}\ {\rm eV}.$

This presentation is based logically on the equality of the absolute values
of the electric charges of the electron and other types of leptons, owing to
which relation (\ref{51}) suggests one more highly important connection:
\begin{equation}
m_{l}^{E}m_{l}^{W}=const.
\label{57}
\end{equation}

Thus, with the availability of the structure of mass, all leptons regardless
of the difference in masses, must have the same charge, an identical charge
radius as well as an equal magnetic moment. Therefore, it is not surprising
that if (\ref{57}) holds, then, for example, leptons universally interact not
only with $\gamma$ or $Z$ but also with any of the corresponding types of gauge
bosons. Of course, such a regularity reflects the characteristic features of
the structure of mass, charge and thereby opens in principle the possibility
for the establishment in nature of universality of all types of interactions
of particles and fields.


\newpage
\begin{thebibliography}{99}
\bibitem{1} O. Klein, Nature {\bf 161}, 897 (1948).
\bibitem{2} E. Clementel and G. Puppi, Nuovo Cimento {\bf 5}, 505 (1948).
\bibitem{3} J. Tiomno, J. Wheeler, Rev. Mod. Phys. {\bf 21}, 144 (1949).
\bibitem{4} T.D. Lee, M.H. Rosenbluth, C.N. Yang, Phys. Rev.
{\bf 75}, 905 (1949).
\bibitem{5} W. Loinaz et. al., Phys. Rev. {\bf D 70}, 113004 (2004)
hep-ph/0403306.
\bibitem{6} W.F. Chang, I.L. Ho, J.N. Ng, Phys. Rev.
{\bf D 66}, 076004 (2002) hep-ph/0203212.
\bibitem{7} A. Masiero, P. Paradisi, R. Petronzio, Phys. Rev. {\bf D 74},
011701 (2006) hep-ph/0511289.
\bibitem{8} P.P. Divakaran, Phys. Rev. {\bf D 60}, 055007 (1999)
hep-th/9903100.
\bibitem{9} M.A. Sanchis-Lozano, Int. J. Mod. Phys. {\bf A 19}, 2183
(2004) hep-ph/0307313.
\bibitem{10} J.H. Park, JHEP, {\bf 10}, 077 (2006) hep-ph/0607280.
\bibitem{11} E. Fermi, Rend. Lincei, {\bf 31}, 184; 306 (1922).
\bibitem{12} E. Fermi, Phys. Zeit. {\bf 23}, 340 (1922).
\bibitem{13} G. Feinberg, L.M. Lederman, Ann. Rev. Nucl.
Sci. {\bf 13}, 157 (1963).
\bibitem{14} R.S. Sharafiddinov, Spacetime Subst. {\bf 3}, 47 (2002)
physics/0305008.
\bibitem{15} S. Weinberg, Phys. Rev. Lett. {\bf 29}, 388 (1972).
\bibitem{16} R.S. Sharafiddinov, Spacetime Subst. {\bf 1}, 176 (2000)
hep-ph/0305009.
\bibitem{17} R.S. Sharafiddinov, Spacetime Subst. {\bf 3}, 86 (2002)
physics/0305009.
\bibitem{18} S.L. Glashow, Nucl. Phys. {\bf 22}, 579 (1961).
\bibitem{19} A. Salam, J.C. Ward, Phys. Lett. {\bf 13}, 168 (1964).
\bibitem{20} S. Weinberg, Phys. Rev. Lett. {\bf 19}, 1264 (1967).
\bibitem{21} R.S. Sharafiddinov, Spacetime Subst. {\bf 3}, 134 (2002)
physics/0305015.
\bibitem{22} R.S. Sharafiddinov, Fizika {\bf B 16}, 1 (2007)
hep-ph/0512346.
\bibitem{23} R.S. Sharafiddinov, in {\it Proceedings of the April Meeting
of the American Physical Society, Dallax, Texas, April 22-25, 2006},
Abstract, H12.00009.
\bibitem{24} R.S. Sharafiddinov, hep-ph/0511065.
\bibitem{25} R.S. Sharafiddinov, Spacetime Subst. {\bf 5}, 83 (2004)
hep-ph/0306255.
\bibitem{26} R.S. Sharafiddinov, Dokl. Akad. Nauk Ruz. Ser. Math.
Tehn. Estest. {\bf 7}, 25 (1998) hep-ph/0307083.
\bibitem{27} R.S. Sharafiddinov, in {\it Proceedings of the April Meeting
of the American Physical Society, Marriott Tampa, Florida, April 16-19, 2005},
Abstract, Z9.00008; hep-ph/0409254.
\bibitem{28} R.S. Sharafiddinov, in {\it Proceedings of the April Meeting
of the American Physical Society, Jacksonville, Florida, April 14-17, 2007},
Abstract, K11.00008.
\bibitem{29} R.S. Sharafiddinov, physics/0702233.
\bibitem{30} R.S. Sharafiddinov, Phys. Essays {\bf 19}, 58 (2006)
hep-ph/0407262.
\bibitem{31} R.B. Begzhanov, R.S. Sharafiddinov, in {\it Proceedings of
the International Conference on Nuclear Physics, Moscow, June 16-19, 1998}
(St-Petersburg, 1998) Abstracts, p. 354.
\bibitem{32} J. Schwinger, Phys. Rev. {\bf 76}, 790 (1949).
\bibitem{33} R.J. Ellis, in {\it Proceedings of the Summer School on Particle
Physics, Zuoz, August 18-24, 2002} (Z\"urich, Switzerland, 2003), p. 1;
hep-ph/0211168.
\bibitem{34} Particle Data Group, {\it Review of Particle Properties},
Phys. Rev. {\bf D 45} (1992).
\end{thebibliography}
\end{document}